\documentclass[usenatbib,useAMS,oneucolumn]{mn2e}

\usepackage{graphicx}
\usepackage[section]{placeins}
\usepackage{mathptmx}
\usepackage{amsmath}

\pagerange{\pageref{firstpage}--\pageref{lastpage}} \pubyear{2014}

\begin{document}

\title[PSR B1133+16 single pulse flux analysis]{Analysis of single pulse radio flux measurements of PSR B1133+16 at 4.85
and 8.35 GHz}

\author[Krzeszowski et. al.]{K.~Krzeszowski,$^1$ O.~Maron,$^1$ A.~S\l{}owikowska,$^1$
J.~Dyks$^2$ and A.~Jessner$^3$ \\
$^1$Kepler Institute of Astronomy, University of Zielona G\'ora, Lubuska 2, 65--265,
Zielona G\'ora, Poland\\
$^2$Nicolaus Copernicus Astronomical Center, Rabia\'nska 8, 87-100 Toru\'n, Poland\\
$^3$Max-Planck-Institut f\"ur Radioastronomie, Auf dem H\"ugel 69, D-53121 Bonn, Germany\\
}
\date{Released 2014}
\maketitle

\begin{abstract}
We show the results of microsecond resolution radio data analysis focused on flux measurements of single pulses of
PSR~B1133+16. The data were recorded at 4.85~GHz and 8.35~GHz with 0.5~GHz and 1.1~GHz bandwidth, respectively, using
Radio Telescope Effelsberg (MPIfR). The most important conclusion of the analysis is, that the strongest single pulse
emission at 4.85~GHz and 8.35~GHz contributes almost exclusively to the trailing part of the leading component of the
pulsar mean profile, whereas studies at lower frequencies report that the contribution is spread almost uniformly
covering all phases of the pulsar mean profile. We also estimate the radio emission heights to be around 1\%--2\% of the
light cylinder radius which is in agreement with previous studies. Additionally these observations allowed us to add two
more measurements of the flux density to the PSR B1133+16 broadband radio spectrum covering frequencies from 16.7 MHz up
to 32 GHz. We fit two different models to the spectrum: the broken power law and the spectrum based on flicker noise
model, which represents the spectrum in a simpler but similarly accurate way.  
\end{abstract}

\label{firstpage}


\section{Introduction}
Pulsar radio emission is still not explained in details but it is believed to originate close to the pulsar surface
(\citealt{krzeszowski2009} and references therein). The analysis of 62 mean profiles of 23 pulsars at different
frequencies regarding aberration and retardation effects yielded the estimation of the emission height being below 1500
km, but most probably being of the order of 500 km above the pulsar surface. The advance in theoretical understanding of
the emission mechanism and conditions in the magnetosphere have been driven mainly by the observations. Recently many
observations have been concentrated on observing single pulses which carry detailed information about the physics of
radio emission. Analysis of single pulse high resolution time series can show a few interesting properties i.e. giant
and bright pulses, subpulse drift, nulling, microstructure, etc. Observations of single pulses for most pulsars may be
carried out mainly at lower frequencies because pulsars are weak radio sources at high frequencies which is clearly seen
in their steep spectra. The pulsar spectrum in general can be described by a power law $S\propto\nu^\alpha$, where
$\alpha$ is the spectral index. The average spectral index for 266 pulsars for frequency spread from 0.4 to 23~GHz is
$\alpha= -1.8$ \citep{mkk+00}.

\begin{table}
\caption{Basic properties of PSR B1133+16 \citep{Brisken2002, manchester2005}.}
\label{tab:props}
    \begin{tabular}{ll}
    \hline 
        BNAME                              &B1133+16\\
        JNAME                              &J1136+1551\\
        $P$                                &1.188 s\\
        $\dot{P}$                          &$3.73 \times 10^{-15}$ s~s$^{-1}$\\
        RA ~~(J2000)                         &11$^\mathrm{h}$36$^\mathrm{m}$03$^\mathrm{s}$\\
        DEC (J2000)                        &15$^\circ$51'04''\\
        DM                                 &4.86 pc~cm$^{-3}$\\
        RM                                 &1.1 rad~m$^2$\\
        Age                                &$5.04 \times 10^6$ Yr\\
        Distance                           &350 $\pm$ 20 pc\\
        Proper motion                       &375 mas yr$^{-1}$\\
        Transverse velocity                &$631^{+38}_{-35}$~km~s$^{-1}$\\
        $B_\mathrm{surf}$                  &$2.13 \times 10^{12}$ G\\
        $\dot{E}$                          &$8.8 \times 10^{31}~\mathrm{erg}~\mathrm{s}^{-1}$\\
     \hline        
    \end{tabular}
\end{table}

PSR B1133+16 is a nearby middle--aged pulsar with one of the highest proper motion, and thus, one
with the highest transverse velocity \citep{Brisken2002}. Basic properties of PSR B1133+16 are gathered in
Table~\ref{tab:props}. Its faint optical counterpart (B=28.1 $\pm$ 0.3 mag) was firstly detected by \cite{Zharikov2008}.
Recently \cite{Zharikov2013} detected the optical candidate of the pulsar counterpart on the GTC and VLT images that
is consistent with the radio coordinates corrected for its proper motion. This source was also detected in X-rays by
\cite{Kargaltsev2006} using the \textit{Chandra} satellite with the flux of (0.8 $\pm$ 0.2) $\times 10^{-14}$
ergs~cm$^{-2}$~s$^{-1}$ in the 0.5--8.0~keV range. For the X--rays fit the assumed hydrogen column density was
$n_\mathrm{H} = 1.5 \times 10^{20}$ cm$^{-2}$. Low value of $n_\mathrm{H}$ and no $H_\alpha$ Balmer bow shock imply a
low density of ambient matter around the pulsar. This pulsar has not been detected by the \textit{Fermi} satellite.

In this paper in Sec. \ref{sec:observations} we describe observational parameters and technical issues about the
recorded data. Sec. \ref{sec:single.pulses} covers analysis of the data. We present two different approaches for data
analysis: mean profiles composed of pulses that their flux fall into specific intensity range as well as the phase
position and flux of single pulses that are stronger than 20$\sigma$. In Sec. \ref{sec:emission.heights} we discuss
radio emission height estimations, while in Sec. \ref{sec:spectrum} we present the radio spectrum of PSR B1133+16 and
discuss different spectrum models. We conclude our results in Sec. \ref{sec:conclusions}.

\section{Observations and data reduction}
\label{sec:observations}
Our analysis is based on the Radio Telescope Effelsberg archival data. The observational settings and parameters are
collected in Table~\ref{tab:params}. Observations were made using the 4.85~GHz and the 8.35~GHz receivers of the MPIfR
100~m radio telescope in Effelsberg. The receivers have circularly polarised feeds. Both receivers feature cryogenically
cooled High-Electron-Mobility Transistor (HEMT) low noise input stages with typical system temperatures of 27~K
for the 4.85~GHz receiver and 22~K for the 8.35~GHz receivers. A calibration signal can be injected synchronously to the
pulse period for accurate measurements of pulsar flux densities. The Effelsberg radio telescope is regularly calibrated
on catalogued continuum sources and we used the mean height of the injected calibration signal of  1.2~K for 4.85~GHz
and 2.083~K for 8.35~GHz for the flux calibrations. The two intermediate frequency (IF) signals (bandwidths of
500~MHz for 4.85~GHz and 1.1~GHz for 8.35~GHz) from each receiver, one for left hand (LHC) and one for right hand (RHC)
circular polarisation are detected in a broad--band polarimeter attached to the receiver, providing four output
independent signals relating to the power levels of LHC, RHC, LHC$\cdot$RHC$\cdot \sin($LHC,RHC$)$, LHC$\cdot$RHC$\cdot
\cos($LHC,RHC$)$. No dedispersion was used before or after the detection due to low value of DM of PSR~B1133+16. The
dispersion broadening amounts to 178 $\mu$s and 78 $\mu$s for 4.85~GHz and 8.35~GHz,
respectively, which is of the order of their sampling rates. The four detected power levels are linearly encoded as
short pulses of a variable frequency of typically 2--3 MHz corresponding to the system noise level, but ranging up to a
maximum of 10 MHz. These signals were brought down to the station building and using the EPOS backend, the four
frequency encoded power levels were recorded synchronously to the pulse period in 1024 phase bins (=samples) per period.
The average signal power for each phase bin was determined by simply counting the number of pulses of the supplied
frequency encoded signal for the duration of the phase bin and then recording the counts on disk for later off-line
processing. Only the data from LHC and RHC channels were used and added together to yield the detected total power for
our analysis. The individual phase bins had durations of 200~$\mu$s for 4.85~GHz and 60~$\mu$s for 8.35~GHz. With the
given system temperatures and the respective antenna efficiencies of 1.5~K/Jy and 1.2~K/Jy, we achieved a typical
sensitivity (rms) of 30~mJy and 80~mJy per single pulse phase bin for the two frequencies, respectively. 

\subsection{Digitisation effects}

The aforementioned sensitivity corresponds to a fraction of $2-3\times 10^{-3}$ of the equivalent background
(baseline) noise level of 16--18~Jy. At the same time, we recorded typically only 600 (4.85~GHz) down to 200 (at
8.35~GHz) frequency encoded counts per phase bin. As a result, one finds that the signal power is resolved in steps of
30~mJy and 80~mJy for the two receivers and that the rms noise fluctuations are barely resolved, amounting to a couple
of counts at most (see Fig.~\ref{fig:correction}).

\subsection{Weather and radio interference}
Weather, especially clouds, changes the opacity of the atmosphere and as a consequence we found that fluctuations of the
sky background were of the order of a fraction of a Kelvin on the short timescales (200~ms and
60~ms) that were used for the measurement window and the subtraction of the noise baseline. Both
receivers were also affected by a very low level 100~Hz modulation ($10^{-3} - 10^{-4}$ of the baseline level)
originating in the receiver's power supply and cooling systems. With typical observed pulse component widths of
5--10~ms, we find that we cannot rule out that weak individual components of individual single pulse may be affected by
the interference. However, averaging and other statistical flux estimates using sufficient numbers ($>$10) of single
pulses would not be affected by weather induced noise fluctuations or power supply interference.

\begin{figure}
\includegraphics[width=84mm]{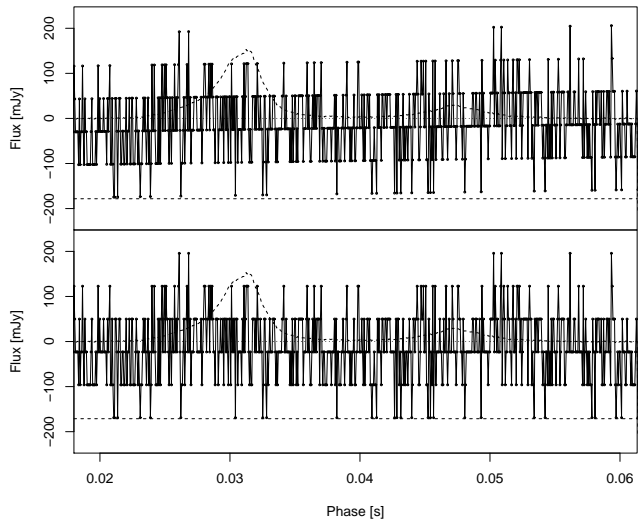}
\caption{Weak single pulse of PSR B1133+16 at 8.35 GHz with a visible linear trend (top panel) and after the
correction (bottom panel). Mean profile (dashed line) is plotted for comparison. Horizontal dashed line is
shown for reference.}
\label{fig:correction}
\end{figure}

The above mentioned effects are visible in the data as approximately linear trends added to each of the recorded single
pulses. One example of such slope and effects of the correction at 8.35~GHz is presented in Fig.~\ref{fig:correction}.
The values of the slopes show no trends from pulse to pulse and they range from -2 to 2 mJy / bin and their
distribution is a Gaussian--like with a zero mean. The behaviour of the system at 4.85~GHz is roughly the same
as of 8.35~GHz data and the same correction routines were applied.

\begin{table}
\caption{Observation parameters.}
\label{tab:params}
\begin{tabular}{lll}
\hline
        Date                               &2002 Feb 07 &2004 Apr 26\\
        Frequency                          &4.85~GHz    &8.35~GHz\\
        Bandwidth                          &0.5~GHz     &1.1~GHz\\
        Observing time                     &67~min.     &120~min.\\
        Number of pulses                   &3361        &6029\\
        Sampling time                      &200~$\mu$s  &60~$\mu$s\\
        Mean flux density                  &1.59~mJy    &0.73~mJy\\
\hline        
\end{tabular}
\end{table}

\section{Single pulses in microsecond resolution}
\label{sec:single.pulses}

\begin{figure}
\includegraphics[width=84mm]{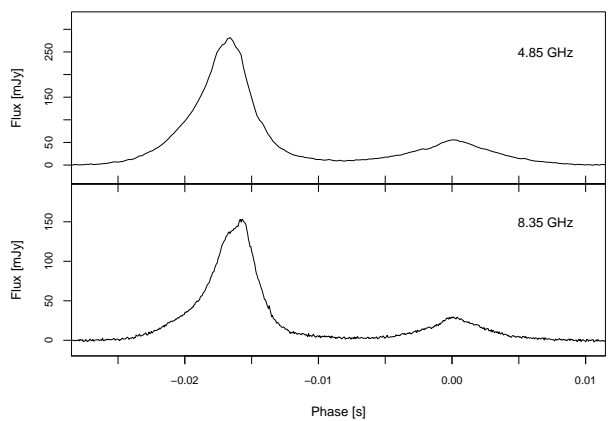}
\caption{Mean profiles of PSR B1133+16 at 4.85~GHz (top panel) and 8.35~GHz (bottom panel). The profiles are aligned
with
respect to the maximum of the trailing component.}
\label{fig:mean.profiles}
\end{figure}

Average profiles of PSR B1133+16 at 4.85~GHz and 8.35~GHz consist of two main components connected by a bridge of
emission (Fig.~\ref{fig:mean.profiles}). The leading component is approximately five times stronger than the trailing
one and they are separated by around 5$^\circ$. The duty cycle of this pulsar is around 3\% at 4.85~GHz and 8.35~GHz. 

\begin{figure}
\includegraphics[width=82mm]{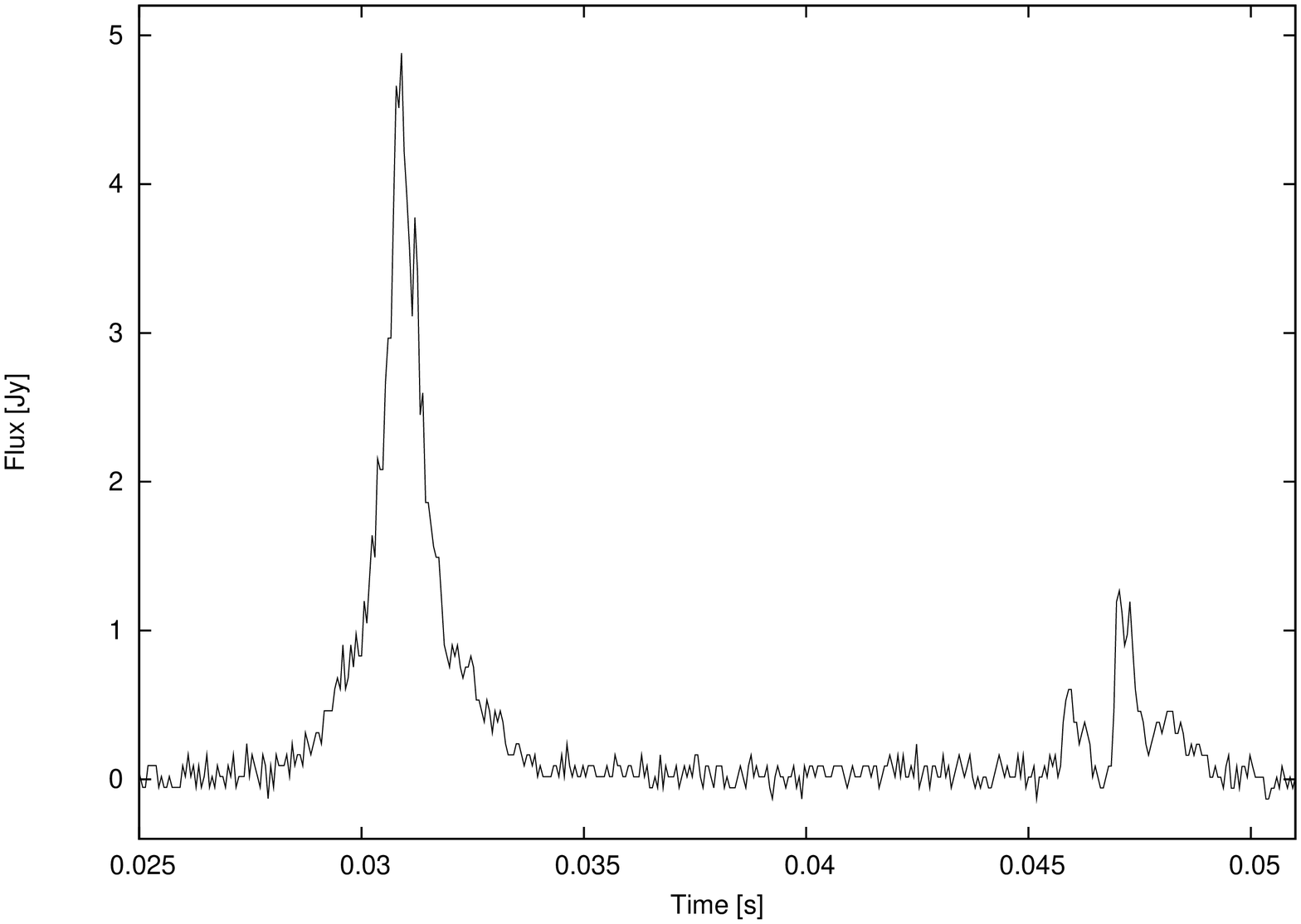}
\includegraphics[width=84mm]{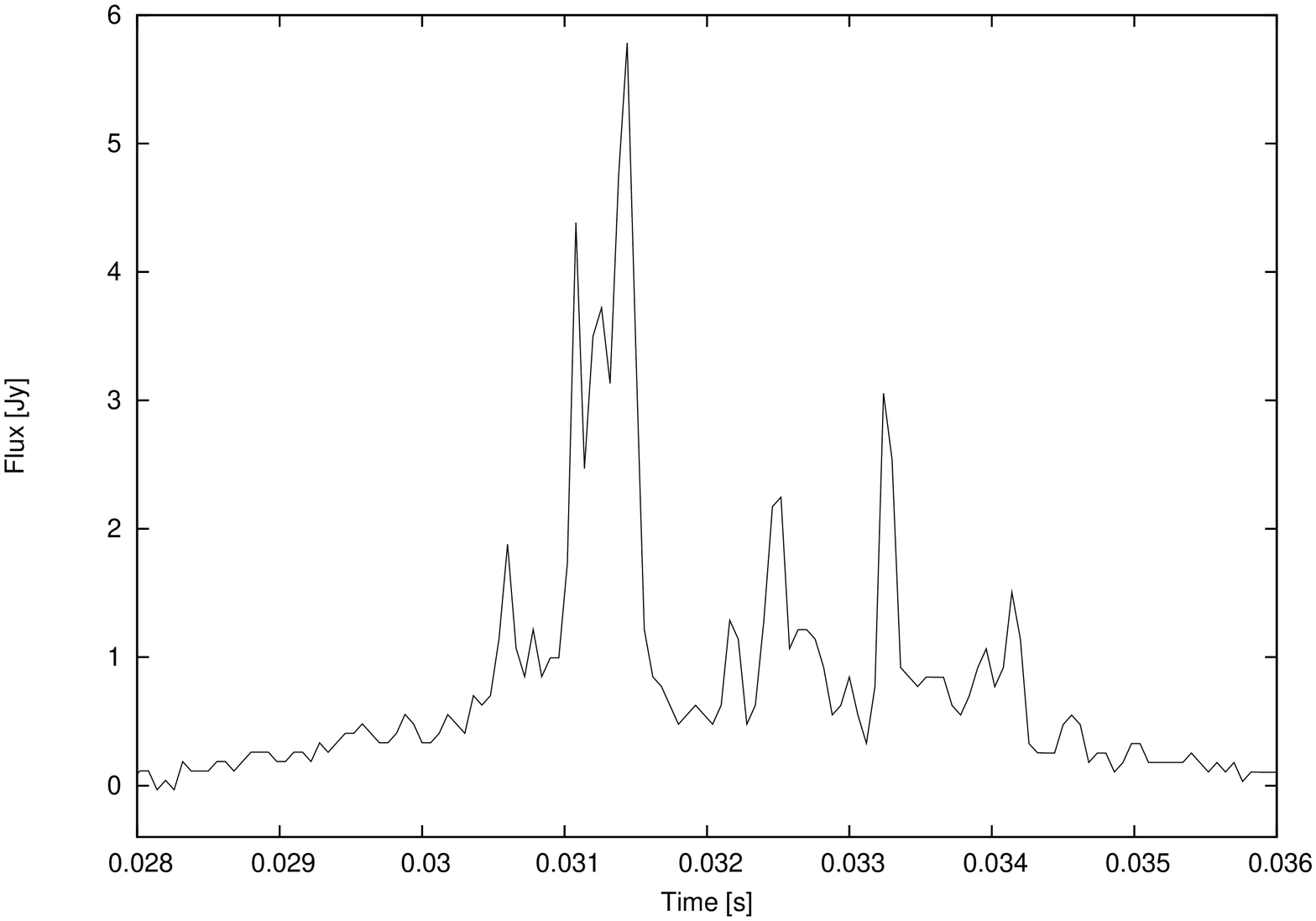}
\caption{An example of a strong single pulse of PSR~B1133+16 at 8.35~GHz with some interesting features in the trailing
component (top panel) and the structure of the leading component of another strong single pulse profile at 8.35~GHz
(bottom panel).}
\label{1133-203}
\end{figure}

We performed an analysis of microsecond resolution radio data focused on flux measurements of single pulses. High
resolution observations allowed us to investigate the microstructure of single pulse shapes. The single pulse at
8.35~GHz in the top panel of Fig.~\ref{1133-203} shows interesting features in the trailing component. On the other
hand the bottom panel of Fig.~\ref{1133-203} is focused on the leading component of another single pulse profile at the
same frequency. As it can be clearly seen, single pulse profiles have complex structures, which can be resolved only
with high time resolution observations. There were numerous studies of the microstructure of PSR~B1133+16 and other
pulsars (e.g. \citealt{ferguson78, lange98}). It is reported that microstructure is most probably related to the
emission process and is present in many pulsars with the fraction of pulses showing microstructure being of the order of
30\% to 70\% \citep{lange98}.

\subsection{Timescale constraints on geometrical parameters}
The timescale of real-time flux variability (such as the micro- or nano--structure) can be interpreted in terms of the
size of the emitting region, or in terms of the angular size of relativistically--beamed radiation pattern
(\citealt{lange98}, \citealt{crossley10}). The outcome depends on the actual spatial and temporal structure of the
emission region, and its relation to the angular scale of the radiated beam.

Let us consider a localised, and relativistically-outflowing source of radio emission, e.g.~a cloud of charges with
Lorentz factor $\gamma$ moving along a \emph{narrow} bunch of magnetic field lines. Here \emph{narrow} means that the
spread of $B$-field direction within the emitting stream, as measured for different rotational azimuths, is smaller than
the intrinsic size of the emission pattern $1/\gamma$. In the observer's reference frame the charges move along a narrow
bunch of trajectories with the radius of curvature $\rho$. If the source emits detectable radiation for a limited period
of time $\Delta t_{\rm em}$ (as measured in our reference frame) an observed spike of radio flux has a width of
$\tau\equiv\Delta t_{\rm obs}= \Delta t_{\rm em}(c-v)/c\simeq \Delta t_{\rm em}/(2\gamma^2)$ (since the outflowing
source is nearly catching up with the emitted photons). If the source persists for a time sufficient to sweep its full
$1/\gamma$ beam across the observer's line of sight, then $\Delta t_{\rm em} \simeq \rho/(\gamma c)$, and the timescale
is:
\begin{equation}
\tau = \tau_{\rho} \simeq \rho/(2 c \gamma^3) = 1.7\times 10^{-9}{\rm s}\ 
\rho_8/\gamma_2^3
\label{taurho}
\end{equation}
where $\rho_8=\rho/10^8$ cm and $\gamma_2 = \gamma/100$ \citep{jackson75}. During that time the point source moves up
in pulsar magnetosphere by a distance of $c\Delta t_{\rm em} \simeq 10^6{\rm cm}\ \rho_8/\gamma_2$. For $\rho=10^8$ cm
the observed timescale of the microstructure of $\tau \sim 100$ $\mu$s limits the Lorentz factor to the mildly
relativistic value $\gamma \simeq 2.6$. For $\gamma=10$ the observed timescale implies $\rho \simeq 6\times 10^{9}$ cm,
comparable to the light cylinder radius of PSR B1133$+$16 ($R_{\rm LC} = 5.67\times 10^9$ cm). Note that for the
curvature radiation from a localised emitter, by definition one expects  $\tau \simeq 1/\nu$, where $\nu$ is the
observed frequency (a few GHz). That is, the pair of values $\rho$ and $\gamma$ must ensure that the curvature spectrum
extends up to the frequency $\nu$. The observed time scale of the microstructure ($\sim\negthinspace10^{-4}$ s) is then
too long to directly correspond to the rapid sweep of the elementary beam of the curvature radiation emitted by a small
plasma cloud.

Another case is encountered when the emitting clouds extend considerably along $B$-field lines ($\Delta x \gg
\rho\gamma^{-3}$), or when there is a steady outflow of uniformly-distributed matter that emits radio waves. Again, let
us first consider the \emph{narrow stream} case, in which the internal spread of rotational azimuths $\Delta \phi_B$ of
$\vec B$ is much smaller than the size of the relativistic beam ($\Delta \phi_B \ll 1/\gamma$). In such a case,
projection of the beam on the sky results in an elongated stripe of width $1/\gamma$. The observed timescale is then
determined by the speed of sightline passage through the $1/\gamma$ stripe, as resulting from the rotation of the
neutron star:  
\begin{equation}
  \begin{aligned}
\tau = \tau_{\rm rot} \simeq P\ (2\pi\gamma\sin\zeta\sin\delta_{\rm cut})^{-1} =  \\
= 1.6\times 10^{-3}P\ (\gamma_2\sin\zeta\sin\delta_{\rm cut})^{-1},
  \end{aligned}
\label{taurot}
\end{equation}
where $\zeta$ is the viewing angle between the sightline and the rotation axis, and $\delta_{\rm cut}$ is the `cut
angle' between the sky-projected emission stripe and the path of the line of sight while it is traversing through the
beam  (see Fig.~2 in \citealt{dyks2012}). \citet{lange98} provide a similar timescale estimate (eq.~6 therein)
which is valid only for $\delta_{\rm cut}=90^\circ$, i.e.~for orthogonal passage of the sightline through the beam. For
the parameters $\alpha=88^\circ$ and $\zeta = 97^\circ$ determined by \citet{ganga99}, the observed separation
of components ($2\phi\simeq5^\circ$) implies $\delta_{\rm cut}=74^\circ$. Eq.~(\ref{taurot}) then gives $\gamma_2 \simeq
20$, which is smaller than the minimum Lorentz factor required for the curvature spectrum to extend
up to the observed frequency of a few GeV.

In a thick stream case, the spread of rotational azimuths of $\vec B$ within the stream is much larger than the angular
size of the stream ($\Delta \phi_B \gg 1/\gamma$). The emitted radiation can be considered (approximately) tangent to
the local magnetic field, and to the charge trajectory in the observer's frame. The observed timescale is then
determined by the angular extent of the stream in the magnetic azimuth $\phi_m$, measured around the dipole axis. For
the stream extending between $\phi_{m,1}$ and $\phi_{m,2}$ the timescale is equal to $\tau=\tau_{\phi m}=\phi_{2} -
\phi_{1}$, where the pulse longitudes $\phi_1(\phi_{m,1})$ and $\phi_2(\phi_{m,2})$ correspond to the moments when our
sightline starts and stops probing the region of the stream. For known (or assumed) $\alpha$, $\zeta$, $\phi_{m,1}$
and $\phi_{m,2}$ the values of $\phi_1$ and $\phi_2$ can be calculated from eqs.~(18) and (19) in \citep{dyks2010}.
However, the timescale of $\tau\sim100$ $\mu$s corresponds to the angle $2\pi\tau/P =  0.03^\circ$ which requires
$\gamma \ge 2\times 10^3$. Thus, for PSR~B1133$+$16 the thick stream case may need to be considered only for $\gamma >$
a few thousands.

\subsection{Single pulse flux distribution}
\begin{figure}
\includegraphics[width=84mm]{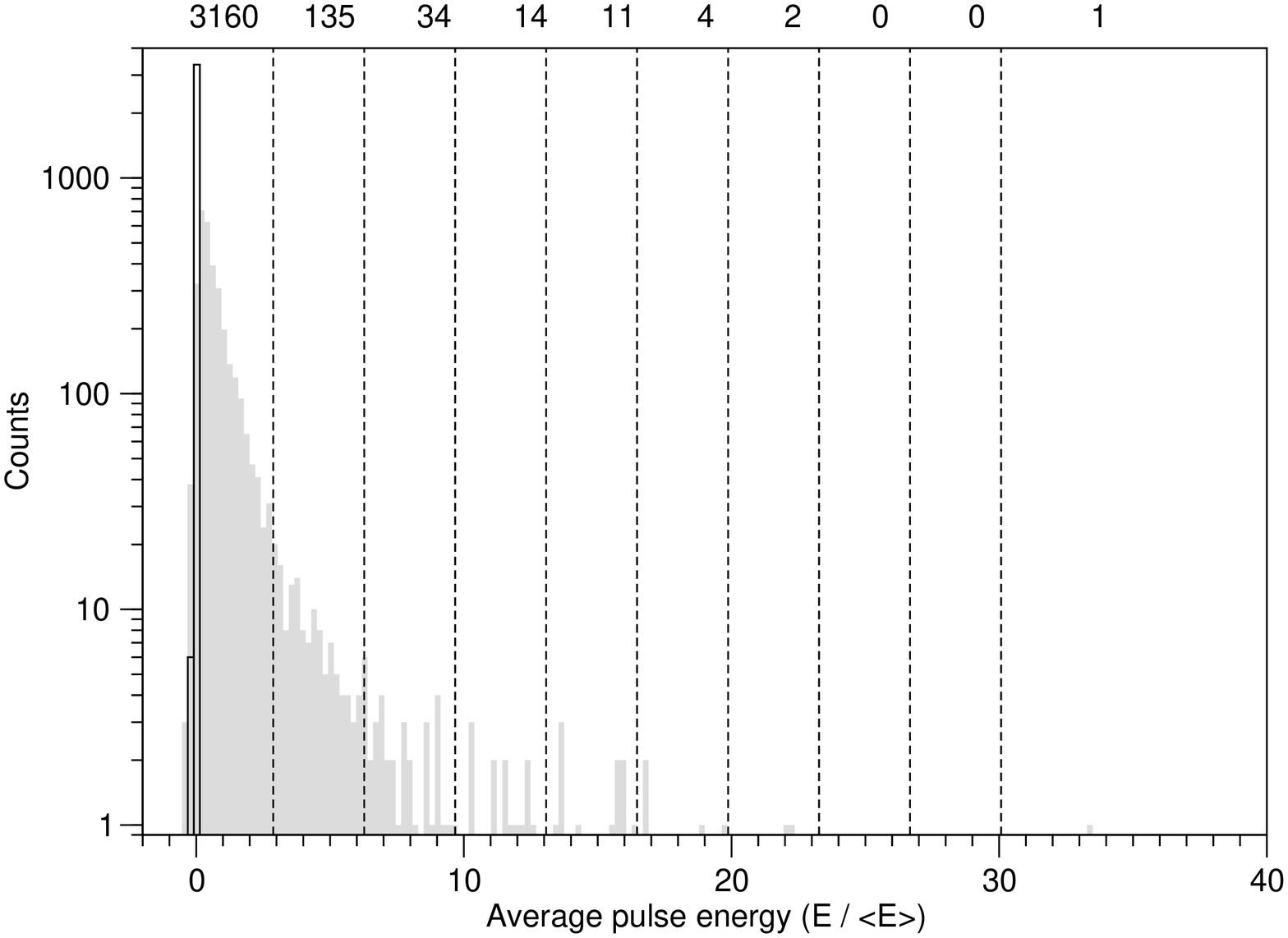}
\includegraphics[width=84mm]{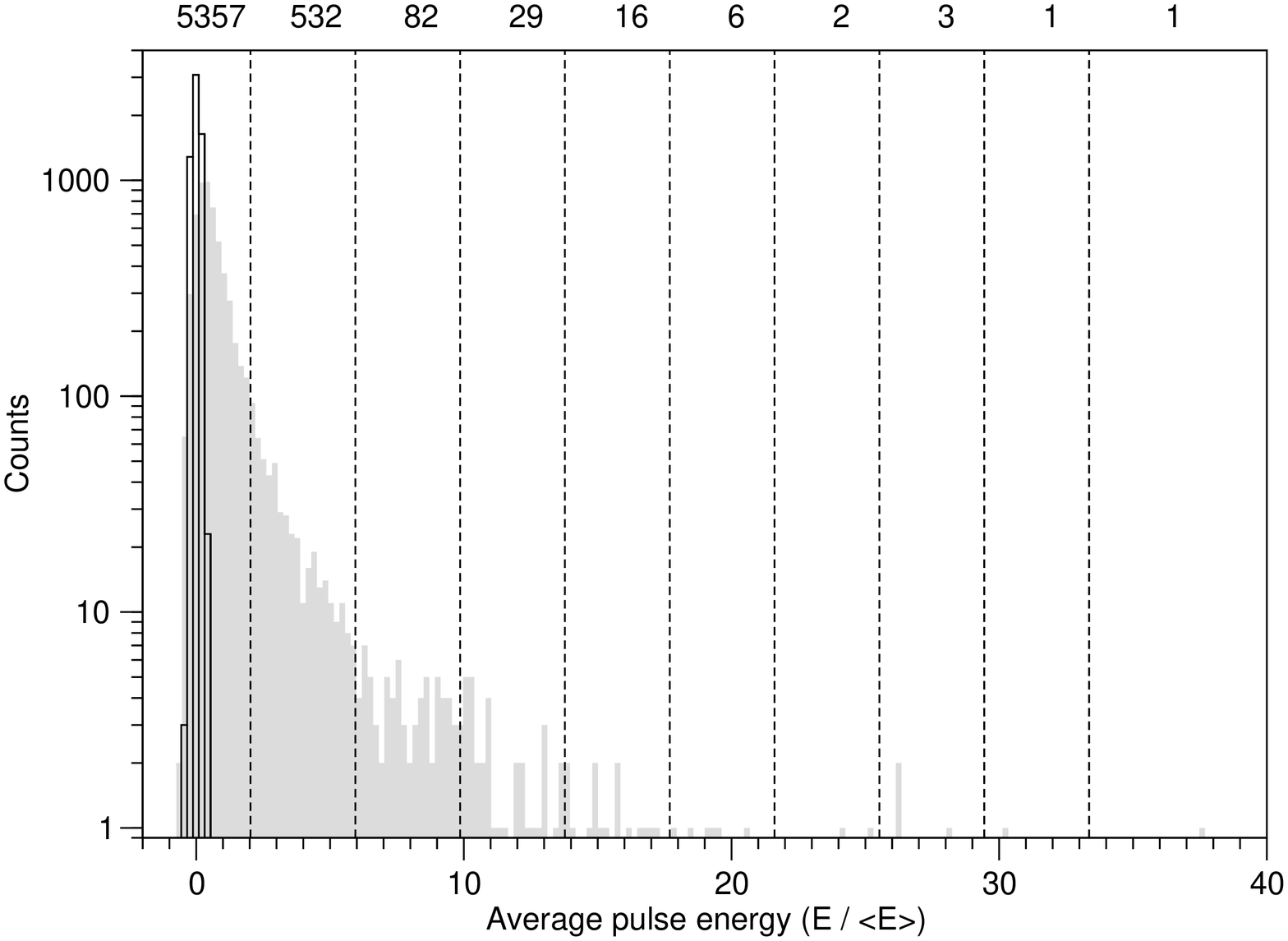}
\caption{Flux density distribution (grey bars) and off--pulse intensity distribution (black outlined bars) of 4.85~GHz
(top panel) and 8.35~GHz (bottom panel) data. Vertical dashed lines
denote ten intensity ranges with a number of pulses that fall into the particular range. Mean profiles composed of
single pulses that fall into a particular range are presented in Fig.~\ref{fig:intensity.ranges} for both frequencies.}
\label{fig:fluxes.histogram}
\end{figure}

We define bright pulses as those with energy of ten times greater than the mean flux of the pulsar. Looking closer at
the flux density distribution (Fig.~\ref{fig:fluxes.histogram}) one can see that there are not many bright pulses in
our time series, only around 0.9\% at both frequencies. 
    
\begin{figure*}
\includegraphics[width=168mm]{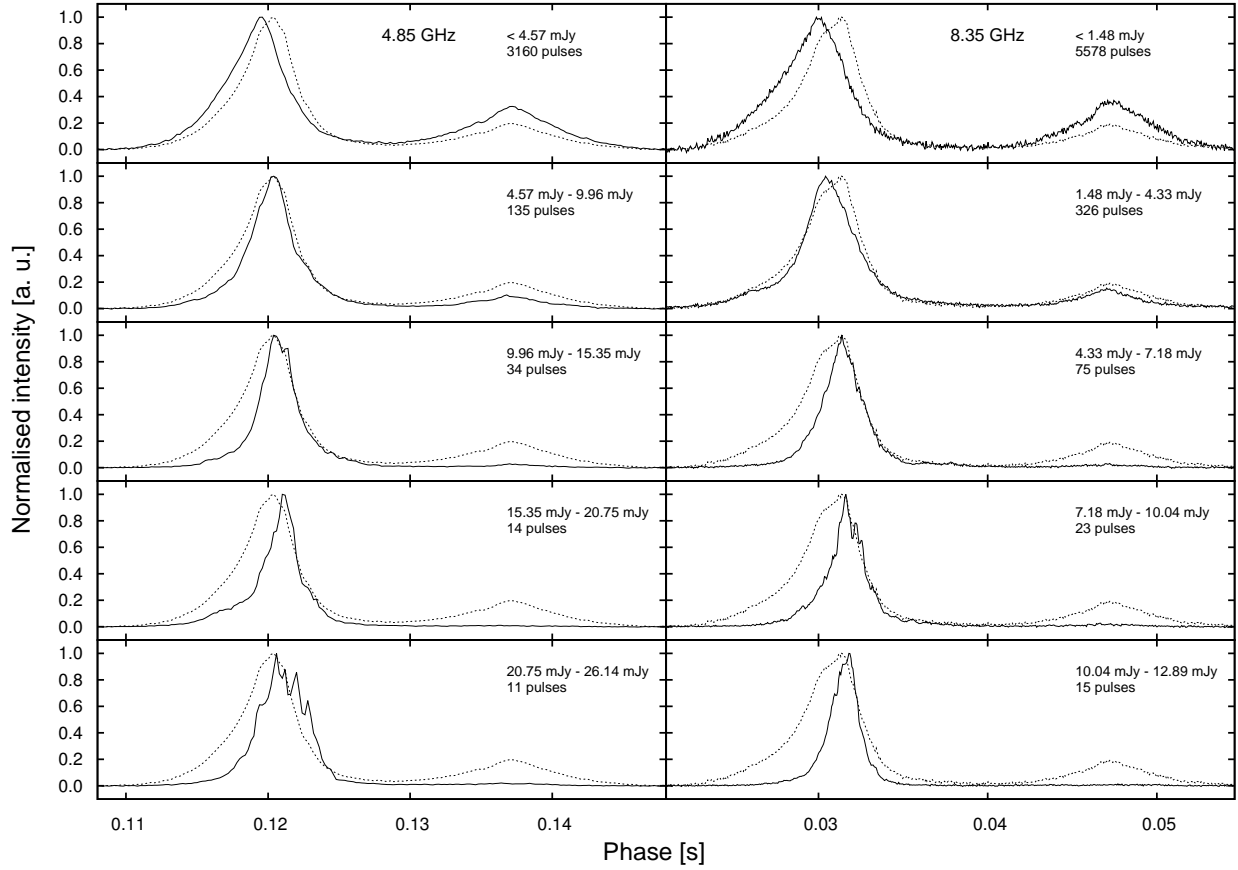}
\caption{Normalised profiles (solid lines) averaged over first five intensity ranges (see
Fig.~\ref{fig:fluxes.histogram}) and the mean profile for all single pulses (dotted line) at 4.85~GHz (left column) and
8.35~GHz (right column).}
\label{fig:intensity.ranges}
\end{figure*}

All flux measurements (Fig.~\ref{fig:fluxes.histogram}) for both frequencies were divided into ten equally sized
intensity ranges \citep{nowakowski96}. The ranges were constructed as follows: the minimum and maximum flux measurements
from each data set were taken as boundaries and all remaining pulse flux values were assigned to one of the
($S_\mathrm{max} - S_\mathrm{min}$) / 10 ranges of the widths of around 5~mJy and 3~mJy for 4.85~GHz and 8.35~GHz,
respectively. The minimum and maximum flux values are $-$0.83~mJy and 53.11~mJy as
well as $-$1.38~mJy and 27.15~mJy for 4.85~GHz and 8.35~GHz data, respectively. Negative flux values are introduced into
the data because of observing system properties which were discussed in detail in Sec.~\ref{sec:observations} and
stochastic noise properties in pulses with no detection. For the single pulses with flux falling into a particular
range the average profiles were constructed. Normalised profiles for 4.85~GHz and 8.35~GHz data and respective mean
profiles of first five intensity ranges are presented in Fig.~\ref{fig:intensity.ranges}. The number of pulses falling
into specific intensity range is written in each panel and also shown in the Fig.~\ref{fig:fluxes.histogram} above the
distributions. In the Fig.~\ref{fig:intensity.ranges} it is easily noticeable that the maximum of the intensity
level averaged profile (solid line) moves towards later phases with respect to the maximum of the mean pulsar profile
(dotted line) for both frequencies and the shift is increasing with frequency from 0.36$^\circ$ at 4.85~GHz to
0.55$^\circ$ at 8.35~GHz. It means that low intensity single pulses contribute mainly to the leading part of
first component, whereas higher intensity single pulses contribute almost exclusively to its trailing part. On the other
hand the second component is composed mainly of the lowest intensity single pulses --- it is visible only in the first
two intensity ranges. The shifts of pulsar components were also investigated by \citet{mitra2007}. They report that the
stronger emission of B0329+54 comes earlier than the weaker emission with a delay of 1.5$^\circ$. This is opposite to
what we have observed. Moreover, Mitra et al. investigated the effect only at one frequency, i.e. 325~MHz. Therefore,
further investigations of more pulsars single pulses are highly recommended because the mechanism behind observed
effect is not understood and one does not know how common it may be in other pulsars.

\begin{figure}
\includegraphics[width=84mm]{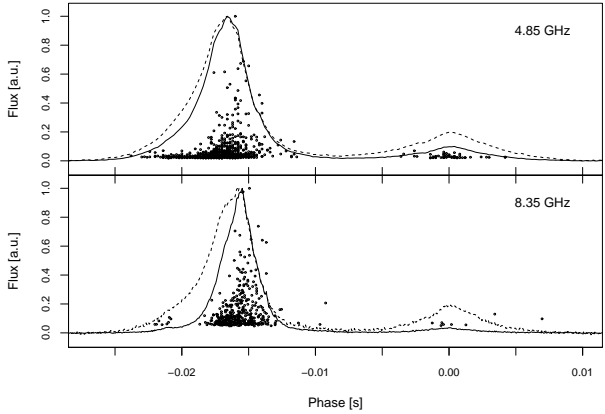}
\caption{Maximum flux positions for single pulses with SNR $> 20 \sigma$ at 4.85~GHz (top panel) and 8.35~GHz (bottom
panel). The mean profile composed of all pulses is plotted with dotted line, whereas the mean profile composed of the
pulses with SNR $> 20 \sigma$ is plotted with solid line. Both mean profiles and flux maxima of individual pulses are
normalised with respect to their maximum values. The profiles are aligned with respect to the maximum of the trailing
component at both frequencies.}
\label{maxima}
\end{figure}

It is clearly visible in Fig.~\ref{maxima} that at 4.85~GHz and 8.35~GHz maximum flux values (denoted with dots)
contribute almost only to the trailing edge of the first component of the mean profile, whereas studies at lower
frequencies report that the contribution is spread almost uniformly covering all phases of the pulse mean profile
\citep{kss+11}. A mean profile composed only of single pulses with SNR~$>~20\sigma$ is plotted with a solid line,
whereas the mean profile composed of all of the single pulses is plotted with a dotted line. The numbers of such pulses
are 758 (23\%) at 4.85~GHz and 407 (7\%) at 8.35~GHz, respectively.

\section{Emission heights}
\label{sec:emission.heights}
In the case of 8.35~GHz data from Fig.~\ref{fig:intensity.ranges} (right hand side panels) one can read
out the shift of 0.0018 seconds ($\Delta\phi~\approx$~0.55$^\circ$~=~0.001~rad) between the longitude of the
low-flux and high-flux emission. Ignoring (for a while) the curved shape of B-field lines, this shift can be translated
to altitude difference: $\Delta r_\mathrm{em}~=~R_\mathrm{LC}~\Delta\phi/2~=~2.7~\times~10^7$~cm (independent of
$\alpha$ and $\zeta$). However, the radius of curvature of B--field lines at the rim of polar cap of this pulsar is
$\rho_\mathrm{B}~=~(4/3)(r_\mathrm{NS}~R_\mathrm{LC})^{1/2}~=~10^8$~cm. The upward shift of emission by
$\Delta r_\mathrm{em}$, results then in a change of emission direction by $\Delta r_\mathrm{em}
/ \rho_\mathrm{B}$~$\approx$~0.27~rad~$\approx$~15$^\circ$, which is much larger than the observed displacement of
0.5$^\circ$. Therefore, if the observed misalignment of low--flux and high--flux emission has anything to do with the
real spatial shift of emission region, it must be dominated by the effect of curved B--field lines rather than the
aberration--retardation shift. Unfortunately, in such a case a full information on geometry is needed to determine
$r_\mathrm{em}$ or $\Delta r_\mathrm{em}$. The same logic applied to 4.85~GHz data
(left hand side panels Fig.~\ref{fig:intensity.ranges}) yields the following results:
$\Delta\phi~\approx~0.36^\circ~=~0.006$~rad, $\Delta
r_\mathrm{em}~\approx~1.8\times10^6$~cm and $\Delta r_\mathrm{em} / \rho_\mathrm{B}$~$\approx~10^\circ$ which is also
much bigger than the measured shift of 0.36$^\circ$.

Using $\alpha = 88^\circ$ and $\beta = 9^\circ$ \citep{ganga99} and canonical formulae \citep{lorimer2005} we derived
radio emission heights at 4.85~GHz and 8.35~GHz. While making the height estimates we associate the peak separation with
two different sets of B--field lines: the last open field lines which have the standard magnetic colatitude
$\sin\theta_\mathrm{pc}=(r_\mathrm{NS}/R_\mathrm{LM})^{1/2}$, and with the critical field lines, which have
$\sin\theta_c=(2/3)^{3/4}(r_\mathrm{NS}/R_\mathrm{LC})^{1/2}$.

Simple calculations yield the estimations of emission heights at 4.85~GHz are 67$\times10^6$cm and 122$\times10^6$cm 
for the last open and critical magnetic field lines, respectively, whereas for stronger emission it is
closer to the neutron star surface at 66$\times10^6$cm
and 120$\times10^6$cm. Similarly, at 8.35~GHz, emission heights are 66$\times10^6$cm and 122$\times10^6$cm for the last
open and critical field lines, respectively, whereas for stronger emission it is 65$\times10^6$cm and 119$\times10^6$cm.
Our analysis shows that the emission region is located at a distance of around 1\%-2\% of the light cylinder radius from
the pulsar surface which is consistent with earlier studies (e.g. \citealt{krzeszowski2009}).

\section{Radio spectrum} 
\label{sec:spectrum}   
\begin{figure*}
\includegraphics[width=168mm]{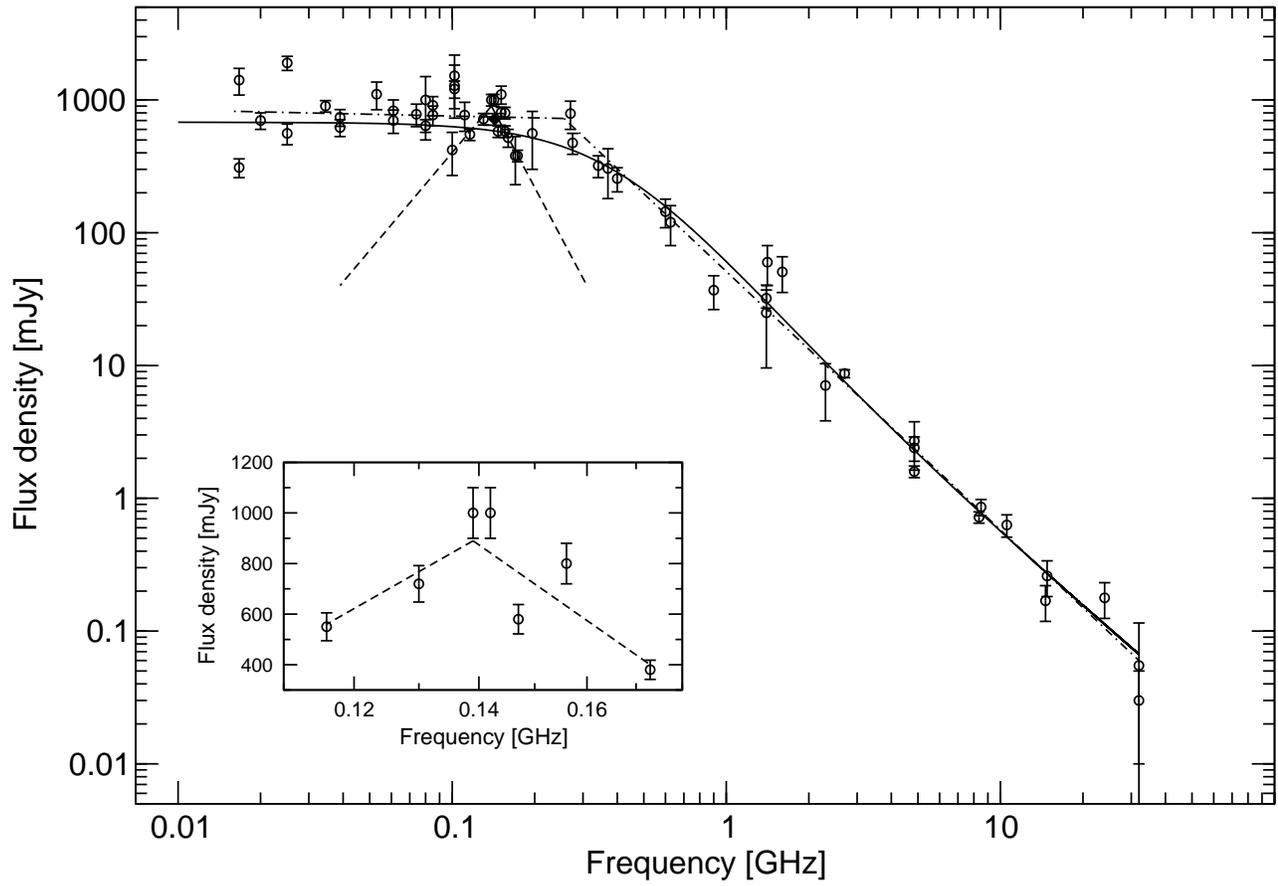}
\caption{Spectrum of PSR B1133+16. Individual measurements are listed in Table~\ref{table:fluxes}. Inset contains
spectrum (dashed line) as presented in \citet{kss+11}. Solid line represents a fitted model using Eq.
\ref{equation:spectrum} \citep{ljk+08}. Dot-and-dash line represents a broken power law fit. See text for details.}
\label{fig:spectrum}
\end{figure*}

In general pulsars have steep spectra with an average spectral index around $-1.8$ \citep{mkk+00}. Except for basic
spectrum (that can be described with power law $S \propto \nu^{\alpha}$) there are two common types: spectrum with a
break (described with two power laws) and spectrum with a turn--over (clearly visible maximum flux). We collected flux
density measurements from different publications (Table~\ref{table:fluxes}). Our dataset covers a very wide radio
frequency range from 16.7~MHz up to 32~GHz. The analysis of the data presented in this paper yielded the mean flux
values of 1.59~mJy and 0.73~mJy at 4.85~GHz and 8.35~GHz, respectively, with an estimated error of 10\% of the original
value. The spectrum of PSR~B1133+16 spanning the wide radio frequency range is shown in Fig.~\ref{fig:spectrum}.
Each point denotes a measurement of mean flux density with respective uncertainties. We included the measurements by
\citet{kss+11} at 7 frequencies, ranging from 116.75~MHz to 173.75~MHz. The authors claim that the spectrum of PSR
B1133+16 over their "reasonably wide frequency range" of 57 MHz is of a broken power law type with spectral indices of
$\alpha_1=2.33\pm2.55$ and $\alpha_2=-3.8\pm2.24$. In Fig.~\ref{fig:spectrum}  we have indicated their spectrum by a
dashed line. We cannot confirm Karuppusamy's spectral indices which were obtained from low frequency measurements
covering only a small range of frequencies. However, we find that Karuppusamy's measurements have a typical spread of
flux values and thereby fit well into the overall spectrum as can be seen in Fig.~\ref{fig:spectrum}. 

Our spectrum of PSR B1133+16 may be described by two power laws in the whole frequency range with
$\alpha_1=-0.04~\pm~0.0001$ and $\alpha_2=-1.96~\pm~0.0001$ with a break frequency of $\nu_b=256~\pm~0.016$~MHz (${\rm
\chi_{red}^2=5.1}$). 
    
To reproduce the spectrum of PSR B1133+16 covering over 32~GHz frequency range we have also fitted the flicker
noise model proposed by \citet{ljk+08} which is described by
\begin{equation}
{\rm S(\omega)=S_0\left(\frac{1+\omega^2\tau_e^2}{\tau_e^2}\right)^{n-1}\!\!\!\!\times e^{-i(n-1)atan(\omega\tau_e)},}
\label{equation:spectrum}
\end{equation}
where S$_0$ is a scaling factor, ${\rm \omega=2\pi\nu}$, $\nu$ is an observing frequency, ${\rm \tau_e}$ is a
characteristic time for the nano--burst decay time, and n is the exponent which constrains a combination of physical
parameters of nano pulses (for details refer to \citealt{ljk+08}). Our fitted parameters of ${\rm
S_0=3.39~\pm~0.77~Jy}$, ${\rm \tau_e=0.40~\pm~0.05~ns}$ and ${\rm n=0.118~\pm~0.022}$ are in good agreement with
\citet{ljk+08} result and our reduced ${\rm \chi_{red}^2=4.7}$ is comparable to the value for the broken power law fit.
The model, apart from the scaling factor S$_0$, is based on only two physical parameters. The first one is the duration
of a nano--pulse ($\tau_\mathrm{e}$) and the second one (n) is related to the geometry of the emission process. It is
based on the assumption that the pulsar radio emission is in fact the superposition of many nano--pulses, which in case
of PSR B1133+16 have duration of $\tau_\mathrm{e} = 0.51$~ns according to \citet{ljk+08}. However our fit gives even
shorter nano--pulse duration of 0.4~ns. \citet{ljk+08} report that for 12 pulsars 0.1~ns~$<~\tau_\mathrm{e}~<$~2.0~ns.

\begin{table*}
\centering
\caption{Flux density measurements for PSR B1133+16 with references. \textbf{References: }[1] - \citealt{mal99}, [2] -
\citealt{izv+81},
  [3] - \citealt{mal+00}, [4] - \citealt{kss+11}, [5] - \citealt{kra+03}, [6] - \citealt{lorimer95}, [7] -
\citealt{mkk+00}, [8] -
\citealt{ljk+08}}
\label{table:fluxes}
  \begin{tabular}{llll@{~~~~~~~~~}llll}
  \hline
Frequency&Flux &Error &Reference&Frequency&Flux &Error &Reference\\
~~[MHz]&[mJy]&[mJy]&&~~[MHz]&[mJy]&[mJy] \\
\hline
16.7    &310        &50     &[1]    &156     &580        &58     &[1]\\
16.7    &1410       &320    &[1]    &160     &520        &80     &[1]\\
20      &700        &100    &[1]    &170     &380        &150    &[1]\\
25      &560        &100    &[1]    &173.75  &380        &38     &[4]\\
25      &1900       &230    &[1]    &196     &560        &260    &[1]\\
34.5    &900        &90     &[1]    &270     &790        &190    &[1]\\
39      &740        &107    &[1]    &275     &475        &85     &[1]\\
39      &620        &90     &[2]    &341     &320        &60     &[5]\\
53      &1105       &260    &[1]    &370     &305        &124    &[1]\\
61      &830        &170    &[1]    &408     &256        &53     &[6]\\
61      &700        &140    &[2]    &606     &144        &35     &[6]\\
74      &780        &150    &[1]    &626     &120        &40     &[5]\\
80      &640        &70     &[1]    &925     &37         &11     &[6]\\
80      &1000       &500    &[1]    &1400    &24.92      &15.33  &[7]\\
85      &910        &150    &[1]    &1408    &32         &5      &[6]\\
85      &770        &130    &[2]    &1412    &60         &20     &[5]\\
100     &420        &150    &[1]    &1606    &51         &15     &[6]\\
102     &1210       &180    &[1]    &2300    &7.07       &3.24   &[7]\\
102     &1280       &550    &[3]    &2700    &8.7        &0.6    &[7]\\
102     &1520       &660    &[1]    &4850    &2.7        &1.07   &[7]\\
102     &1020       &200    &[2]    &4850    &2.4        &0.5    &[5]\\
111     &770        &190    &[1]    &4850    &1.59       &0.16   &this paper\\
116.75  &550        &55     &[4]    &8350    &0.73       &0.07   &this paper\\
130     &720        &72     &[4]    &8500    &0.86       &0.12   &[7]\\
139.75  &1000       &100    &[4]    &10550   &0.63       &0.12   &[7]\\
142.25  &1000       &100    &[4]    &14600   &0.169      &0.0507 &[7]\\
147.5   &580        &58     &[4]    &14800   &0.26       &0.078  &[7]\\
151     &1100       &170    &[1]    &24000   &0.178      &0.0534 &[7]\\
151     &805        &75     &[1]    &32000   &0.03       &0.02   &[7]\\
156     &800        &80     &[4]    &32000   &0.055      &0.06   &[8]\\
\hline

\end{tabular}
\end{table*}
    
\section{Conclusions}
\label{sec:conclusions}

The analysis of PSR B1133+16 single pulses is a process that needs a certain amount of care. First off all, it is
important to take into account different observational and technical effects that can affect recorded data, especially
with very high time resolution. The effects that are presented in this paper play huge role and alter the data
significantly. Understanding of such effects and their influence on the data recording process is important for proper
data reduction. Some of the effects are not visible in the mean profiles but only in single pulses data.

The mean profiles of PSR B1133+16 at 4.85~GHz and 8.35~GHz consist of two components (Fig.~\ref{fig:mean.profiles}). The
second component is emitted almost exclusively by low intensity individual pulses. On the other hand, the first
component is seen in single pulses regardless of their intensity except for the cases when it is not present at all.
However, lower intensity emission contributes mostly to the leading part of the first component whereas higher intensity
single pulses contribute mainly to its trailing part (Fig.~\ref{fig:intensity.ranges}) which was also reported by
\citet{maron2013}. The results of analysis of 4.85~GHz and 8.35~GHz data are consistent with previous studies by
\citep{nowakowski96} at 430~MHz but studies of B0329+54 \citep{mitra2007} show entirely opposite behaviour without full
explanation. This inconsistency requires further studies of other pulsars single pulses to explain this effect.

We show, in contradiction to studies at lower frequencies by
\citet{kss+11}, who report an almost uniform spread
of single pulse maxima, that the maximum emission of B1133+16
single pulses at 4.85 and 8.35~GHz contributes almost exclusively
to the trailing part of the leading component of the mean profile.
Our result is consistent with the behaviour at 341, 626, 1412 and
4850~HMz mentioned by \citet{kra+03} and extends the
studies up to 8.35 GHz.

Radio emission arises close to the pulsar surface at the distances of around 65 stellar radii at frequencies of 4.85~GHz
and 8.35~GHz. Weaker emission, which contributes to the leading part of the leading components, comes in earlier phases
which suggests that originates in magnetosphere further from the pulsar surface than more energetic emission. Our
calculations shows, that the difference of the emission heights for stronger and weaker emission is of the order of a
few stellar radii, amounting to a change of 1\%--2\% of the emission height, which is consistent with previous
estimations \citep{krzeszowski2009}.

There are 60 mean flux measurements in the literature of PSR B1133+16 that are known to us. They span a very wide radio
frequency range from 16.7~MHz up to 32~GHz. To reproduce the spectrum we fitted two different models: the broken
power--law model and one based on flicker noise model of pulsar radio emission \citep{ljk+08}. Surprisingly, the model
proposed by \citet{ljk+08} is not widely used in the literature although it reproduces the pulsar spectrum
comparably well to the power-law model. Future high time resolution observations might be useful to verify nano--pulse
emission model. 

Due to the fact, that the pulsar radio emission is weaker at higher frequencies, giant pulses are the ones that can
allow us to study closely their structure. In our both data samples there is roughly one per cent of bright pulses that
are at least ten times stronger than mean flux and their microstructure is clearly visible. 

\section*{Acknowledgements}
The presented results are based on the observations with the 100--m telescope of the 
MPIfR (Max-Planck-Institut f\"ur Radioastronomie) at Effelsberg. This work has been supported by Polish
National Science Centre grants DEC-2011/03/D/ST9/00656 (KK, AS), DEC-2012/05/B/ST9/03924 (OM) and
DEC-2011/02/A/ST9/00256 (JD). Data analysis and figures were partly prepared using R \citep{rcite}.

\footnotesize{
  \bibliographystyle{mn2e}
  \bibliography{krzeszowski}

\begin{thebibliography}{}

\bibitem[\protect\citeauthoryear{{Brisken}, {Benson}, {Goss} \&
  {Thorsett}}{{Brisken} et~al.}{2002}]{Brisken2002}
{Brisken} W.~F.,  {Benson} J.~M.,  {Goss} W.~M.,    {Thorsett} S.~E.,  2002,
  ApJ, 571

\bibitem[\protect\citeauthoryear{{Crossley}, {Eilek}, {Hankins} \&
  {Kern}}{{Crossley} et~al.}{2010}]{crossley10}
{Crossley} J.~H.,  {Eilek} J.~A.,  {Hankins} T.~H.,    {Kern} J.~S.,  2010,
  ApJ, 722

\bibitem[\protect\citeauthoryear{{Dyks} \& {Rudak}}{{Dyks} \&
  {Rudak}}{2012}]{dyks2012}
{Dyks} J.,  {Rudak} B.,  2012, MNRAS, 420

\bibitem[\protect\citeauthoryear{{Dyks}, {Rudak} \& {Demorest}}{{Dyks}
  et~al.}{2010}]{dyks2010}
{Dyks} J.,  {Rudak} B.,    {Demorest} P.,  2010, MNRAS, 401

\bibitem[\protect\citeauthoryear{{Ferguson} \& {Seiradakis}}{{Ferguson} \&
  {Seiradakis}}{1978}]{ferguson78}
{Ferguson} D.~C.,  {Seiradakis} J.~H.,  1978, A\&A, 64

\bibitem[\protect\citeauthoryear{{Gangadhara}, {Xilouris}, {von Hoensbroech},
  {Kramer}, {Jessner} \& {Wielebinski}}{{Gangadhara} et~al.}{1999}]{ganga99}
{Gangadhara} R.~T.,  {Xilouris} K.~M.,  {von Hoensbroech} A.,  {Kramer} M.,
  {Jessner} A.,    {Wielebinski} R.,  1999, A\&A, 342

\bibitem[\protect\citeauthoryear{{Izvekova}, {Kuzmin}, {Malofeev} \&
  {Shitov}}{{Izvekova} et~al.}{1981}]{izv+81}
{Izvekova} V.~A.,  {Kuzmin} A.~D.,  {Malofeev} V.~M.,    {Shitov} I.~P.,  1981,
  A\&AS, 78

\bibitem[\protect\citeauthoryear{{Jackson}}{{Jackson}}{1975}]{jackson75}
{Jackson} J.~D.,  1975, {Classical electrodynamics}.
Wiley

\bibitem[\protect\citeauthoryear{{Kargaltsev}, {Pavlov} \&
  {Garmire}}{{Kargaltsev} et~al.}{2006}]{Kargaltsev2006}
{Kargaltsev} O.,  {Pavlov} G.~G.,    {Garmire} G.~P.,  2006, ApJ, 636

\bibitem[\protect\citeauthoryear{{Karuppusamy}, {Stappers} \&
  {Serylak}}{{Karuppusamy} et~al.}{2011}]{kss+11}
{Karuppusamy} R.,  {Stappers} B.~W.,    {Serylak} M.,  2011, A\&A, 525

\bibitem[\protect\citeauthoryear{{Kramer}, {Karastergiou}, {Gupta}, {Johnston},
  {Bhat} \& {Lyne}}{{Kramer} et~al.}{2003}]{kra+03}
{Kramer} M.,  {Karastergiou} A.,  {Gupta} Y.,  {Johnston} S.,  {Bhat} N.~D.~R.,
     {Lyne} A.~G.,  2003, A\&A, 407

\bibitem[\protect\citeauthoryear{{Krzeszowski}, {Mitra}, {Gupta}, {Kijak},
  {Gil} \& {Acharyya}}{{Krzeszowski} et~al.}{2009}]{krzeszowski2009}
{Krzeszowski} K.,  {Mitra} D.,  {Gupta} Y.,  {Kijak} J.,  {Gil} J.,
  {Acharyya} A.,  2009, MNRAS, 393

\bibitem[\protect\citeauthoryear{{Lange}, {Kramer}, {Wielebinski} \&
  {Jessner}}{{Lange} et~al.}{1998}]{lange98}
{Lange} C.,  {Kramer} M.,  {Wielebinski} R.,    {Jessner} A.,  1998, A\&A, 332

\bibitem[\protect\citeauthoryear{{L{\"o}hmer}, {Jessner}, {Kramer},
  {Wielebinski} \& {Maron}}{{L{\"o}hmer} et~al.}{2008}]{ljk+08}
{L{\"o}hmer} O.,  {Jessner} A.,  {Kramer} M.,  {Wielebinski} R.,    {Maron} O.,
   2008, A\&A, 480

\bibitem[\protect\citeauthoryear{Lorimer}{Lorimer}{2005}]{lorimer2005}
Lorimer D.,  2005, {Handbook of Pulsar Astronomy}.
Cambridge Observing Handbooks for Research Astronomers, Cambridge University
  Press

\bibitem[\protect\citeauthoryear{{Lorimer}, {Yates}, {Lyne} \&
  {Gould}}{{Lorimer} et~al.}{1995}]{lorimer95}
{Lorimer} D.~R.,  {Yates} J.~A.,  {Lyne} A.~G.,    {Gould} D.~M.,  1995, MNRAS,
  273

\bibitem[\protect\citeauthoryear{Malofeev}{Malofeev}{1999}]{mal99}
Malofeev V.~M.,  1999, Katalog radiospektrov pul'sarov, Pushchino:PRAO

\bibitem[\protect\citeauthoryear{{Malofeev}, {Malov} \&
  {Shchegoleva}}{{Malofeev} et~al.}{2000}]{mal+00}
{Malofeev} V.~M.,  {Malov} O.~I.,    {Shchegoleva} N.~V.,  2000, Astronomy
  Reports, 44

\bibitem[\protect\citeauthoryear{{Manchester}, {Hobbs}, {Teoh} \&
  {Hobbs}}{{Manchester} et~al.}{2005}]{manchester2005}
{Manchester} R.~N.,  {Hobbs} G.~B.,  {Teoh} A.,    {Hobbs} M.,  2005, AJ, 129

\bibitem[\protect\citeauthoryear{{Maron}, {Kijak}, {Kramer} \&
  {Wielebinski}}{{Maron} et~al.}{2000}]{mkk+00}
{Maron} O.,  {Kijak} J.,  {Kramer} M.,    {Wielebinski} R.,  2000, A\&AS, 147

\bibitem[\protect\citeauthoryear{{Maron}, {Serylak}, {Kijak}, {Krzeszowski},
  {Mitra} \& {Jessner}}{{Maron} et~al.}{2013}]{maron2013}
{Maron} O.,  {Serylak} M.,  {Kijak} J.,  {Krzeszowski} K.,  {Mitra} D.,
  {Jessner} A.,  2013, A\&A, 555

\bibitem[\protect\citeauthoryear{{Mitra}, {Rankin} \& {Gupta}}{{Mitra}
  et~al.}{2007}]{mitra2007}
{Mitra} D.,  {Rankin} J.~M.,    {Gupta} Y.,  2007, MNRAS, 379

\bibitem[\protect\citeauthoryear{{Nowakowski}}{{Nowakowski}}{1996}]{nowakowski%
96}
{Nowakowski} L.~A.,  1996, ApJ, 457

\bibitem[\protect\citeauthoryear{{R Core Team}}{{R Core Team}}{2013}]{rcite}
{R Core Team} 2013, R: A Language and Environment for Statistical Computing.
R Foundation for Statistical Computing, Vienna, Austria

\bibitem[\protect\citeauthoryear{{Zharikov} \& {Mignani}}{{Zharikov} \&
  {Mignani}}{2013}]{Zharikov2013}
{Zharikov} S.,  {Mignani} R.~P.,  2013, MNRAS, 435

\bibitem[\protect\citeauthoryear{{Zharikov}, {Shibanov}, {Mennickent} \&
  {Komarova}}{{Zharikov} et~al.}{2008}]{Zharikov2008}
{Zharikov} S.~V.,  {Shibanov} Y.~A.,  {Mennickent} R.~E.,    {Komarova} V.~N.,
  2008, A\&A, 479

\end{thebibliography}
}

\bsp
\label{lastpage}
\end{document}